# Studies of Muon-Induced Radioactivity at NuMI

David Boehnlein for the JASMIN Collaboration

*D. J. Boehnlein, A. F. Leveling, N. V. Mokhov\*, K. Vaziri – Fermi National Accelerator Laboratory*
*Y. Iwamoto, Y. Kasugai, N. Matsuda, H. Nakashima\*, Y. Sakamoto - Japan Atomic Energy Agency*
*M. Hagiwara, Hiroshi Iwase, N. Kinoshita, H. Matsumura, T. Sanami, A. Toyoda - High Energy Accelerator Research Organization (KEK)*
*H. Yashima - Kyoto University Research Reactor Institute*
*H. Arakawa, N. Shigyo - Kyushu University*
*H. S. Lee - Pohang Accelerator Laboratory*
*K. Oishi - Shimizu Corporation*
*T. Nakamura - Tohoku University*
*Noriaki Nakao - Aurora, Illinois*
*\* - spokesperson*

**Abstract.** The JASMIN Collaboration has studied the production of radionuclides by muons in the muon alcoves of the NuMI beamline at Fermilab. Samples of aluminum and copper are exposed to the muon field and counted on HpGe detectors when removed to determine their content of radioactive isotopes. We compare the results to MARS simulations and discuss the radiological implications for neutrino factories and muon colliders.

## INTRODUCTION

JASMIN is the Japanese-American Study of Muon Interactions and Neutron Detection, designated by Fermilab as experiment T972. Its purpose is to undertake a study of shielding and radiation physics effects at high-energy accelerators. The experimental goals of JASMIN include benchmarking Monte Carlo codes, improvements to radiation safety, and a study of muon interactions with matter. An understanding of such interactions is important in considering the activation of accelerator components and shielding requirements at an accelerator that produces muons in very large quantities. We present here the status of work in progress to study radioactivation of materials in the muon alcoves of the NuMI neutrino beamline at Fermilab.

## THE NUMI MUON ALCOVES

The NuMI neutrino beam at Fermilab was designed to provide an intense muon-neutrino beam for the MINOS neutrino oscillation experiment, with detectors at Fermilab and the Soudan Underground Laboratory in Soudan, Minnesota. Details of the NuMI beamline are published elsewhere.[1] Briefly, protons from the Fermilab Main Injector are directed onto a graphite target, producing π mesons which are focused by a set of magnetic horns. The target and horns can be configured to deliver a neutrino beam of low, medium, or high energy. The mesons focused by the horns travel through a decay region, where neutrinos are produced via the reaction $\pi^+ \rightarrow \mu^+ + \nu_\mu$. A beam dump absorbs undecayed mesons and leftover primary protons, while the $\mu^+$ and $\nu_\mu$ proceed into the dolomite rock underlying Fermilab between the hadron absorber and the MINOS Near Detector Hall. A series of alcoves are excavated into the rock to monitor the muons and, indirectly, the neutrino beam. To accomplish this, the 3 upstream alcoves each contain a muon monitor comprising an array of ionization chambers to measure the distribution and relative intensity of the muon distribution for each beam pulse. During this study, the Main Injector delivered pulses of approximately $2.5 \times 10^{13}$ protons every 1.9 seconds. Since virtually all particles other than muons and neutrinos are filtered out by the intervening rock, the NuMI alcoves are an ideal location to study muon interactions.

## Layout of the NuMI Alcoves

The NuMI beamline includes 4 muon alcoves. One of these is in the Absorber Hall and the other 3 are penetrations into the dolomite downstream. The layout is shown in Figure 1.

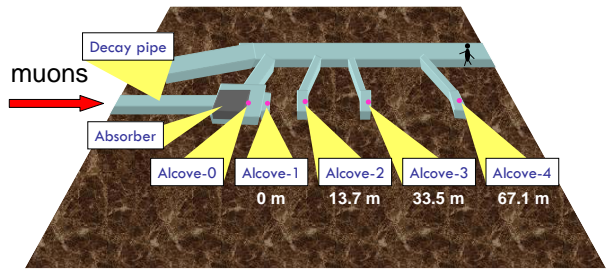

**FIGURE 1.** Layout of the NuMI muon alcoves

**TABLE 1.** Estimated Muon Flux from Monte Carlo; fluence is cm$^{-2}$ 10$^{-12}$ protons/pulse. (from Kopp, *et al.*)

| Location | Charged Particle Fluence | Beam Size |
|---|---|---|
| Alcove 1 | 6.5 × 10$^5$ | 190 cm |
| Alcove 2 | 0.9 × 10$^5$ | 250 cm |
| Alcove 3 | 0.35 × 10$^5$ | 190 cm |

Table 1 shows the calculated charged particle fluence in the first three alcoves.[2]

The estimates in Table 1 assume a low-energy configuration for the NuMI beam. The beam size indicated is FWHM. The calculations also indicate that the neutron content of the fluence in the downstream alcoves (2-4) is < 1%. It is important to note that Alcove 1 is located at the downstream end of the NuMI Absorber Hall and thus may be exposed to any neutrons produced in the NuMI hadron absorber.

## Activation Processes

Radioactivity can be induced in material exposed to a muon beam. As shown in Figure 2, nuclear interactions can be caused by photon exchange between the muon and nucleus. This can produce neutrons, which can produce further nuclear reactions. Either of these processes can produce an unstable, i.e., radioactive nucleus. This process is not ordinarily considered a source of radiological hazard, owing to the weakness of the photo-nuclear interactions relative to the hadronic interactions that dominate the activation of materials at a high energy accelerator.

The JASMIN group has undertaken a series of measurements to assess the radioactivation of materials by muons by placing samples in the NuMI alcoves, where muons are the predominant form of radiation.

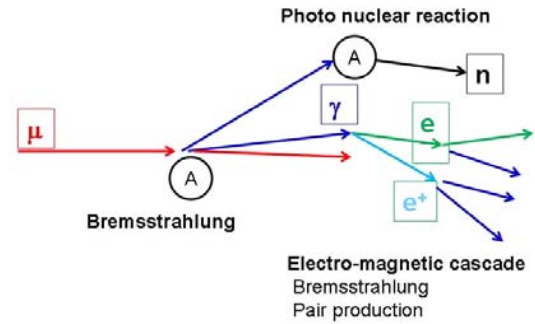

**FIGURE 2.** Photo-nuclear reactions from muons.

## EXPERIMENTAL PROCEDURE

Aluminum and copper were chosen as the materials for the muon activation tests. The samples were disks of 8-cm diameter and 1-cm thickness. The chemical purity of the samples was 99.999% (Al) and 99.994% (Cu). They were placed in the alcoves 1 – 4 and positioned downstream of the muon monitors at approximate beam center. Additional activation tags of aluminum, gold and indium were placed in the alcoves to measure neutron activation. The samples remained in place for 22.8 hours, during which the NuMI beam operated in the low-energy configuration. During this exposure, the NuMI beam delivered 6.26 × 10$^{17}$ protons on target. The duration of the exposure was determined by operational beam stoppages, which allowed access to the alcoves. When removed from the alcoves, the samples were immediately taken to a counting lab set up for this purpose on the Fermilab site, where the samples were placed on High-Purity Germanium counters to study the isotopic abundances of gamma-emitting radionuclides.

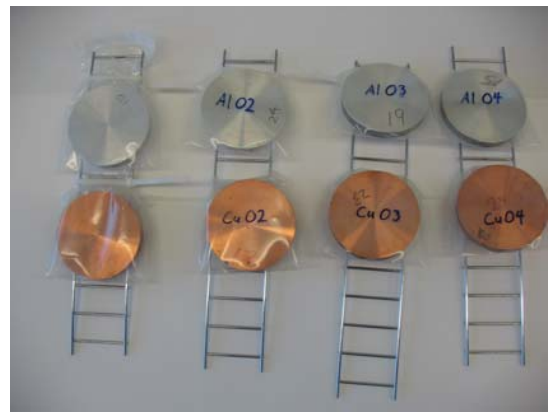

**FIGURE 3.** JASMIN activation samples

# PRELIMINARY RESULTS

A summary of the yield ratios between alcoves is shown in Figure 4. The yield ratio, normalized to the samples in Alcove 2, is plotted against the number of nucleons emitted from the target nucleus. The aluminum target is assumed to contain 27 nucleons; the copper target 63.5 nucleons. The Al target yielded $^{24}$Na and the Cu target yielded isotopes ranging from $^{24}$Na to $^{64}$Cu.

Figures 5 and 6 show the absolute mass yields for copper samples in Alcoves 1 and 2, respectively. Note that the Alcove 1 samples yield $^{24}$Na and $^{64}$Cu, possibly produced by neutron capture, whereas these radionuclides are absent from the samples in Alcove 2.

Additional evidence for neutron interactions in Alcove 1 is shown in Figure 7. This plot shows the attenuation of the yield ratios as a function of depth in the rock downstream of the hadron absorber.[3] As in Figure 4, the numbers are normalized to the yield ratio in Alcove 2. The yield ratios in Alcove 1 are considerably higher than would be expected by extrapolating the results from the other alcoves. The exact composition of the particle flux in Alcove 1 remains to be determined.

The narrow line is a fit to an empirical formula by Rudstam[4] for production by photo-spallation. The process for spallation by fast muons is seen to be similar. The heavy line in Figure 6 represents the result of a Monte Carlo simulation using MARS15[5]. We find the MARS prediction agrees reasonably well with the data, although Monte Carlo underestimates the higher mass number production by roughly a factor of 2. Improvements are being made to MARS15 and additional simulations are planned.

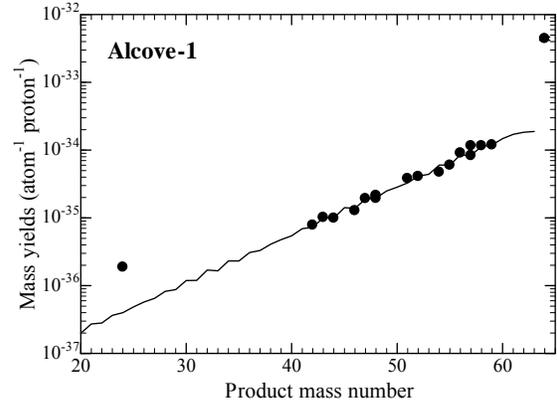

**FIGURE 5.** Activation product mass yields in Alcove 1, per target atom per proton on target. The line is a fit to Rudstam's formula.

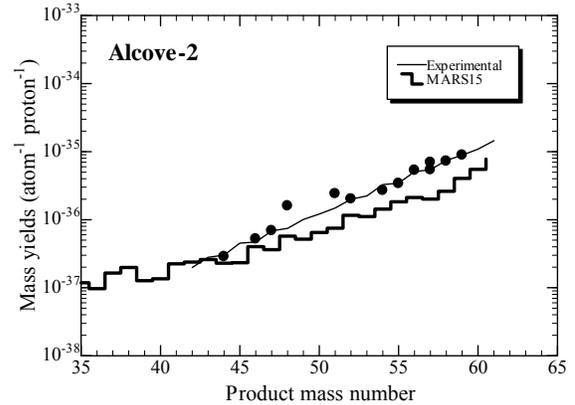

**FIGURE 6.** Activation product mass yields in Alcove 1, per target atom per proton on target. The narrow line is a fit to Rudstam's formula; the heavy line is a calculation by MARS15.

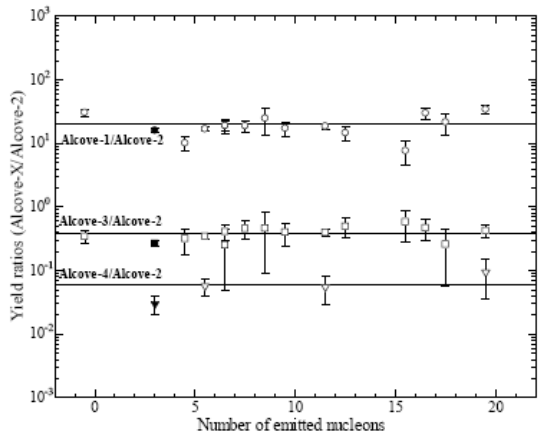

**FIGURE 4.** Yield ratios of activation products normalized to measurements in Alcove 2. The closed and open symbols denote the experimental values from Al and Cu targets, respectively. The solid lines indicate the average of the experimental values.

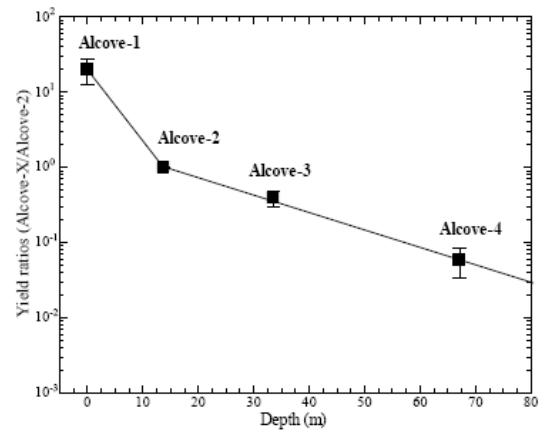

**FIGURE 7.** Attenuation of radionuclide production in the NuMI muon alcoves. The yield ratios are normalized to Alcove 2.

## IMPLICATIONS FOR A MUON ACCELERATOR

The above results are preliminary in the sense that it remains unclear how much of the activation of materials is due directly to muon-induced photo-spallation and how much is due to neutrons or other hadrons produced by muon interactions in the rock walls of the alcoves. However, if one is primarily interested in the overall activation of accelerator components in an operational setting, this might be only a minor consideration, since muons striking such components will presumably generate hadrons in a manner similar to those that interact in the rock wall.

Historically, radioactivation due to muons has not been a serious consideration in the design of particle accelerators. Even in the NuMI muon alcoves, one of the most intense sources of muon radiation in existence, the aluminum framework of the muon monitors is not measurably radioactive with standard survey instruments, such as the Bicron log survey meter. Typical primary beam intensities for NuMI were $2 \times 10^{13}$ protons per pulse during this exposure. Applying this intensity to the entries in Table 1, one would expect muon intensities of approximately $13 \times 10^6$ cm$^{-2}$ per pulse in Alcove 1.

However, the muon intensities are expected to be much higher at a neutrino factory and/or muon collider. Such a machine is anticipated to be designed for acceleration of muons on the order of $10^{21}$ per year.[6] About a half of those muons decay during stores, giving rise to intense electromagnetic showers with copious production of energetic (up to hundreds GeV in a muon collider) Bethe-Heitler muons. A percent of a primary most energetic muon beam will be catched in a beam collimation system. Both of these sources will result in muon fluxes several orders of magnitude higher than the muon intensity in the NuMI alcoves. The possibility that, under these circumstances, muons might produce significant levels of radioactivity in accelerator components and their environs should be taken into account in the design and operational planning of such a machine.

## ACKNOWLEDGMENTS

This work has been supported by grand-aid of ministry of education (KAKENHI 19360432) in Japan. Fermilab is a U.S. Department of Energy Laboratory operated under Contract DE-AC02-07CH11359 by the Fermi Research Alliance, LLC. Yoshimi Kasugai, Hiroshi Matsumura and Toshiya Sanami have provided the figures used here.